\documentclass{webofc}
\usepackage[varg]{txfonts}   
\begin{document}
\title{Quantum to classical parton evolution in the QGP}
%
%

\newcommand{\beq}{\begin{eqnarray}}
\newcommand{\eeq}{\end{eqnarray}}
\newcommand{\dis}{\displaystyle}
\def\dd{{\rm d}}
\def\ii{{\mathrm i}}
\newcommand{\bem}{\begin{multline}}
\newcommand{\eem}{\end{multline}}
\newcommand{\beg}{\begin{gather}}
\newcommand{\eeg}{\end{gather}}
\newcommand{\noi}{\noindent}
\newcommand{\nn}{\nonumber\\}
\newcommand{\dgg}{^{\dagger}}
\def\fig#1{{Fig.~\ref{#1}}}
\newcommand{\ben}{\begin{eqnarray*}}
\newcommand{\een}{\end{eqnarray*}}
\newcommand{\un}[1]{\underline{#1}}

\newcommand{\eq}[1]{\begin{align}#1\end{align}}
\setlength{\parskip}{1mm}
\def\Ket#1{\left\| #1\right\rangle }
\def\Bra#1{\left\langle #1\right\| }

\newcommand{\eqn}[1]{Eq.~\eqref{#1}}
\newcommand{\fign}[1]{Fig.~\ref{#1}}

\def\cG{{\cal G}}
\def\cP{{\cal P}}
\def\cK{{\cal K}}
\def\cM{{\cal M}}
\def\cD{{\cal D}}
\def\cT{{\cal T}}
\def\cU{{\cal U}}
\def\cF{{\cal F}}
\def\cJ{{\cal J}}
\def\cZ{{\cal Z}}
\def\cH{{\cal H}}
\def\cN{{\cal N}}
\def\cO{{\cal O}}
\def\cS{{\cal S}}
\def\cQ{{\cal Q}}
\def\cW{{\cal W}}

\def\lg{{\langle}}
\def\rg{{\rangle}}
\def\bs{\boldsymbol }
\def\beps{{\boldsymbol \epsilon}}

\def\bdel{\bs\partial}
\newcommand{\secn}[1]{Section~1}
\newcommand{\appn}[1]{Appendix~1}
\newcommand{\sst}{\scriptscriptstyle }
\long\def\comment#1{ }

\def\br{\text{br}}
\def\BH{\text{BH}}
\def\el{\text{el}}
\def\inel{\text{inel}}
\def\med{\text{med}}
\def\Tr{\text{Tr}}

\def\Tjet{\Theta_\text{jet}}
\def\reco{\text{reco}}
\def\corr{\text{corr}}
\def\bkg{\text{bkg}}
\def\jet{\text{jet}}
\def\tot{\text{tot}}
\def\raw{\text{raw}}
\def\sig{\text{sig}}
\def\tru{\text{truth}}

\def\and{\quad\text{and}\quad}
\def\with{\quad\text{with}\quad}
\def\for{\quad\text{for}\quad}
\def\cut{\text{cut}}

\newcommand{\dif}{{\rm d}}
\newcommand{\rmd}{{\rm d}}
\newcommand{\rmdt}{{\rm d} t}
\newcommand{\rme}{{\rm e}}
\newcommand{\rmi}{i}
\newcommand{\rmy}{{y}}
\newcommand{\rmtr}{{\rm tr}}
\newcommand{\rmTr}{{\rm Tr}}
\newcommand{\rmH}{{\rm H}}
\newcommand{\rmI}{{\rm I}}
\newcommand{\rmJ}{{\rm J}}
\newcommand{\rmK}{{\rm K}}
\newcommand{\rmR}{{\rm Re}}

\def\bp{{p}}
\def\bk{{ k}}
\def\bq{{ q}}

\def\bell{{\boldsymbol \ell}}
\def\q{{\boldsymbol q}}
\def\0{{\boldsymbol 0}}
\def\p{{\boldsymbol p}}
\def\l{{\boldsymbol l}}
\def\k{{\boldsymbol k}}
\def\m{{\boldsymbol m}}
\def\n{{\boldsymbol n}}
\def\0{{\boldsymbol 0}}
\def\x{{\boldsymbol x}}
\def\y{{\boldsymbol y}}
\def\X{{\boldsymbol X}}
\def\Y{{\boldsymbol Y}}
\def\D{{\boldsymbol D}}
\def\r{{\boldsymbol r}}
\def\s{{\boldsymbol s}}
\def\z{{\boldsymbol z}}
\def\u{{\boldsymbol u}}
\def\v{{\boldsymbol v}}
\def\w{{\boldsymbol w}}
\def\b{{\boldsymbol b}}
\def\Q{{\boldsymbol Q}}
\def\P{{\boldsymbol P}}
\def\M{{\boldsymbol M}}
\def\K{{\boldsymbol K}}
\def\R{{\boldsymbol R}}
\def\T{{\boldsymbol T}}
\def\btheta{{\boldsymbol \theta}}

\def\bkappa{{\boldsymbol \kappa}}
\def\bbkappa{\bar{\boldsymbol \kappa}}
\def\bnu{{\boldsymbol \nu}}
\def\V{\hat{\boldsymbol v}_{1}}
\def\bV{\hat{\boldsymbol v}_{2}}
\def\bK{\hat{\boldsymbol k}_{2}}
\def\qqb{{q\bar q}}
\def\sM{\text{med}}
\def\Qs{ l_\perp^{-1}}
\def\td{ t_\text{decoh}}
\def\ttaufull{\kappa}
\def\ttau{\kappa}
\def\max{{\rm max}}

\newcommand{\del}{\partial}
\newcommand{\lap}{\nabla_^2}
\newcommand{\lan}{\langle}
\newcommand{\ran}{\rangle}
\newcommand{\avg}[1]{\langle #1 \rangle}
\newcommand{\order}[1]{\mcal{O}{(#1)}}
\newcommand{\mcal}{\mathcal}
\newcommand{\pr}{\mcal{P}}
\newcommand{\ob}{\mcal{O}}
\newcommand{\E}{\mcal{E}}
\newcommand{\F}{\mcal{F}}
\newcommand{\lb}{\lambda}
\newcommand{\wt}{\widetilde}

\newcommand{\bmp}{\bm{p}}
\newcommand{\bmk}{\bm{k}}
\newcommand{\bmx}{\bm{x}}
\newcommand{\bmy}{\bm{y}}
\newcommand{\bmu}{\bm{u}}
\newcommand{\bmv}{\bm{v}}
\newcommand{\bmz}{\bm{z}}
\newcommand{\bmr}{\bm{r}}
\newcommand{\bmw}{\bm{w}}
\newcommand{\bmeps}{\bm{\varepsilon}}

\newcommand{\tp}{\acute{t}}
\newcommand{\xp}{\acute{x}}
\newcommand{\zp}{\acute{z}}
\newcommand{\rp}{\acute{r}}

\newcommand{\barg}{\bar{\gamma}}
\newcommand{\atpi}{\frac{\bar{\alpha}}{2 \pi}}
\newcommand{\abar}{\bar{\alpha}}

\author{\firstname{João} \lastname{Barata}\inst{1}\fnsep\thanks{email: jlourenco@bnl.gov} \and
        \firstname{Jean-Paul} \lastname{Blaizot}\inst{2}\fnsep\and
        \firstname{Yacine} \lastname{Mehtar-Tani}\inst{1}\fnsep
}

\institute{Physics Department, Brookhaven National Laboratory, Upton, NY 11973, USA
\and
           Institut de Physique Théorique, Université Paris Saclay, CEA, CNRS, F-91191 Gif-sur-Yvette, France
          }

\abstract{%

  We study the time evolution of the density matrix of a high energy quark in the presence of a dense QCD background that is modeled as a stochastic Gaussian color field. At late times, we find that only the color singlet component of the quark's reduced density matrix survives the in-medium evolution and that the density matrix becomes asymptotically diagonal in both transverse position and momentum spaces. In addition, we observe that the quantum and classical quark entropies converge at late times. We further observe that the quark state loses all memory of the initial condition. Combined with the fact that the reduced density matrix satisfies Boltzmann-diffusion transport, we conclude that the quark reduced density matrix can be interpreted as a classical phase space distribution. 
}
\maketitle
\section{Introduction}
\label{intro} 
Jets offer an optimal tool to explore the multiple stages involved in the time evolution of hot QCD matter created in heavy ion collisions. To that end, several jet observables have been studied over recent years, capturing both global and local properties of the fragmentation cascade in the medium. In parallel, a lot of progress has been achieved in the description of medium induced modifications to the jet structure due to the interactions with the matter. One important question, which we follow to address at the level of single parton evolution, is the extent to which the medium effects are \textit{truly} quantum, and can not be consistently captured in a classical description. These aspects can, in general, be addressed by analyzing the jet density matrix. Moreover, this description allows to directly compute interesting observables such as the jet entropy. Below, we show an exploratory calculation of this quantity in the  of heavy ion collisions  and argue that it can be highly sensitive to the medium scales.

\section{Single parton density matrix evolution in a QGP }
\begin{figure}
    \centering
    \includegraphics[width=.5\textwidth]{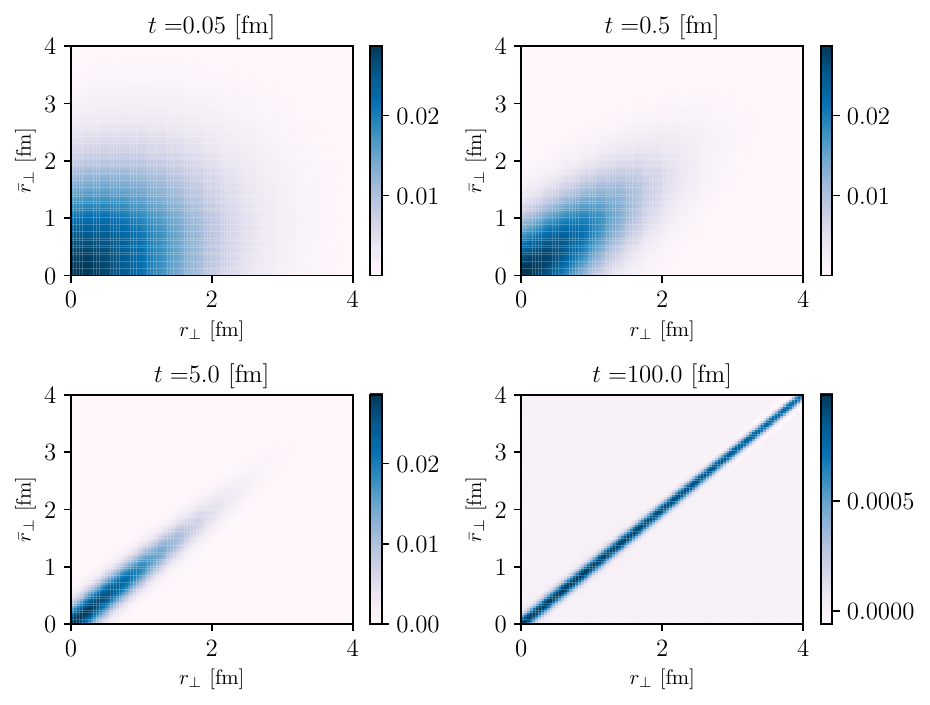}
    \caption{Position space evolution of the reduced density matrix~\cite{Barata:2023uoi}.}
    \label{fig:1}
\end{figure}

We consider first the single parton case, where the jet evolution is dominated by the dynamics of the leading parton. We consider that this particle evolves in the presence of a dense quark gluon plasma (QGP), which can be represented in terms of a stochastic gauge field with statistics: 
\begin{align}\label{eq:2-pt-correlator}
 g^2\Big \langle A^{a} (\q,t)A^{\dagger b} ( \q',t') \Big \rangle_{\!A} &= \delta^{ab } \delta(t-t') (2\pi)^2 \delta^{(2)}(\q-\q')\, \gamma(\q) \, .
\end{align}
In this case the parton's (here a quark) reduced density matrix is obtained after tracing over all the medium's degrees of freedom
\begin{align}\label{rhoArho}
\rho \equiv {\rm tr}_{\!_A} \!\left( \rho[A] \right)= \Big\langle  |\psi_A(t) \rangle \langle \psi_A(t) |  \Big\rangle_{\!A}\,,
\end{align}
where $|\psi_A(t) \rangle$ represents the state of the leading parton for a given field configuration at time $t$. This density matrix has projections in both singlet and octet color subspaces. The octet term, in the absence of a coherent magnetic color field, is quickly damped and only the singlet survives the time evolution in the matter~\cite{Barata:2023uoi}. The evolved density matrix satisfies the following Boltzmann transport  
 \begin{align}\label{eq:rho_s-mom}
  \langle \k | \rho_{\rm s}(t)| \bar \k\rangle&=C_F  \int_{\q}  \int_0^t dt' \, e^{i\frac{(\k^2-\bar \k^2)}{2E}(t-t')}   \gamma(\q)\left[   \langle \k-\q | \rho_{\rm s}(t')| \bar \k-\q\rangle     -  \langle \k | \rho_{\rm s}(t')| \bar \k\rangle   \right]\,,
\end{align}
with $\gamma\sim q^{-4}$ the in-medium elastic scattering rate. Working in the small angle scattering limit, Eq.~\eqref{eq:rho_s-mom} reduces to a diffusion equation, describing the parton's momentum diffusion in the medium. Notice that in this case, intermediate kinetic phases are taken into account, which usually drop out when computing the momentum broadening distribution. In Fig.~\ref{fig:1} we show the time evolution of the density matrix $\rho$ in a position representation as a function of time. Starting from a Gaussian wave-packet with non-zero width, we observe that the time evolution leads first to a spreading along the diagonal. This mechanism becomes relevant once the momentum exchanges with the medium become of the order of the typical initial momentum scale. At late time, the density matrix becomes highly diagonal. This indicates that the full object can be properly described starting from a classical statistical approach. 

\section{Single parton entropy in the medium }
\begin{figure}
    \centering
    \includegraphics[width=.45\textwidth]{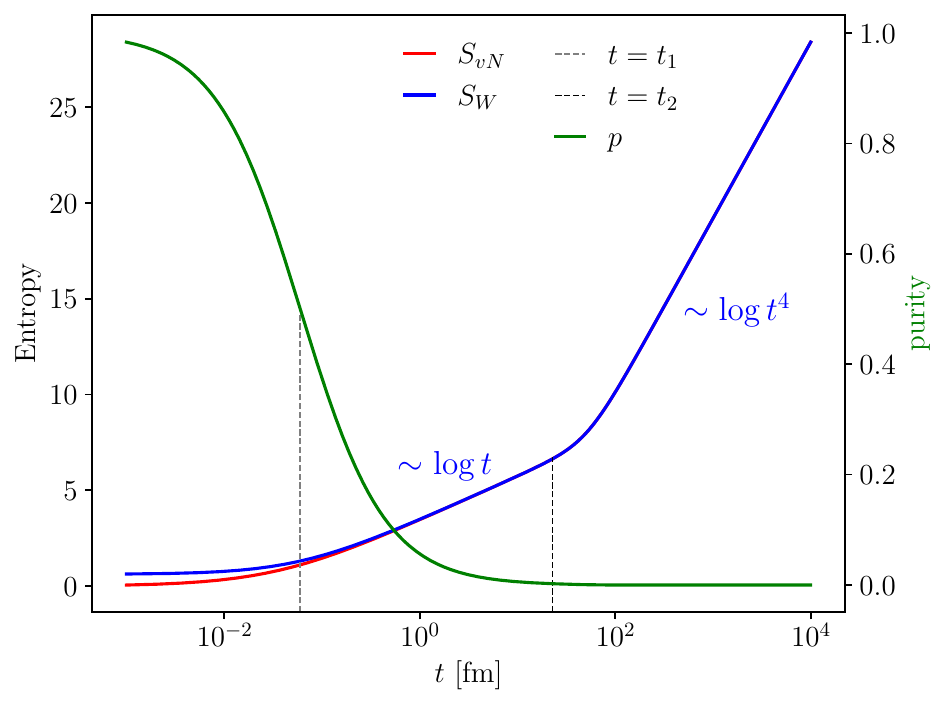}
    \caption{Time evolution of the entrpy and state purity, with characteristic scalings being indicated~\cite{Barata:2023uoi}.}
    \label{fig:2}
\end{figure}

Having computed the density matrix for a single parton in the medium, we follow to see if the diagonalization of $\rho$ is captured at the level of the associated entropy. To that end we compute the von-Neumann entropy
\begin{align}\label{eq:vN-entropy-def}
 S_{\rm vN}[\rho] =-  \Tr  \rho \ln  \rho =  \log \left(\frac{1-p}{4p}\right) + \frac{1}{\sqrt{p}}\,  \ln \frac{1+p +2 \sqrt{p}}{(1-p)} \, ,
\end{align}
which for the current problem can be fully expressed in terms of the purity $p =\Tr \rho^2$. To establish the presence of true quantum features, we introduce the (classical) Wigner entropy
\begin{align}\label{eq:W-entropy}
S_{_{\rm W}} \equiv - \int_{\K,\b} \rho_{_{\rm W}} (\b,\K) \log  \rho_{_{\rm W} }(\b,\K) \, .  
\end{align}
At asymptotic times, one can show that $\frac{S_{_{\rm W}}-S_{\rm vN}} {S_{_{\rm W}}}  \approx \frac{\sqrt{p}}{\ln(1/p)} $, and the two entropies can be identified. Furthermore one can show that in this regime the entropy grows exactly with the logarithm of the phase space, as expected for a classical distribution. The combination of all these facts supports the interpretation of the parton density matrix as a classical distribution. In Fig.~\ref{fig:2} we show the time evolution of the entropy and purity. At late times, the entropy grows in an accelerated fashion, once spatial and momentum diffusion are possible. The collapse of the Wigner and von-Neumann entropies occurs after scatterings with the medium become sizable~\cite{Barata:2023uoi}.

\section{Medium modified jet entropy }
Having discussed the case of a single parton, we now consider the evolution of the entropy for a full jet at leading logarithmic accuracy, following~\cite{Neill:2018uqw}. Here we consider the modifications to the jet entropy due to i) energy loss effects, ii) medium induced gluon radiation. To that end, we introduce the von-Neumann entropy measured on the hardest subjets, having a minimal energy fraction $z_c$ and anguar resolution $R_c$: 
\begin{align}
   \cS = -\sum_n \int d\Pi_n \frac{dP}{d\Pi_n} \log \frac{dP}{d\Pi_n} \, ,
\end{align}
where $\Pi_n$ denotes the phase for $n$ subject and $dP_n$ the probability to produce such a state. 

To include energy loss, we use the quenching weight approximation~\cite{Baier:2001yt} to account for a global quench of the $n$ particle cross-section. It is instructive to consider the one parton limit, where one directly has 
\begin{align}
 \cS = - \int_\q \frac{dP_0}{d^2\q} Q \log \left(\frac{dP_0}{d^2\q} Q\right) = \cS_0 \, Q + \cS(Q)\, ,
\end{align}
where $Q<1$ denotes the one-body quenching factor, $\cS_0$ denotes the single parton entropy in the vacuum, and $\cS(Q)$ is the entropy gained by the fact that energy has been lost. Thus, in this simple case, the entropy gets both a damping term and an additional piece, which competes with the quenching factor. At $\mathcal{O}(\alpha_s)$, a similar result is found:
\begin{align}
  \cS_Q^{\mathcal{O}(\alpha_s)}(E,R)&= \frac{2\alpha_s}{\pi} \Bigg(Q_g \log \frac{1}{z_c} \log \frac{R}{R_c} + \cS(Q^{(2)})  + \int_{z_c}^1 \frac{dz}{z} \int_{R_c}^R \frac{d\theta}{\theta} Q_0^{(2)}(z,\theta) \log \frac{8\pi \alpha_s \Lambda^2}{z^2 \theta^2 E^2}\Bigg) \, ,
\end{align}
where $\Lambda$ is dimensionful regulator. Again, we see a quenching factor multiplying the single parton entropy, which competes with the entropy related to a two body energy loss process. 

The entropy can also be computed in the limit where energy loss is neglected, but the production of induced radiation is taken into account. Focusing on the leading order modification, we use that the in-medium cross-section can be decomposed as $\frac{d\sigma}{\sigma} = \left(1+F_{\rm med}\right) \frac{d\sigma_{\rm vac}}{\sigma} $, to write the entropy variation $\Delta S = S - S_{\rm vac}$ approximately as
\begin{align}
  R\frac{d[\Delta \cS^{\alpha_s}](E)}{dR} \approx \frac{2\alpha_s}{\pi } \Bigg\{ \int_{z_c}^1 \frac{dz}{z}    \log \left[\frac{e^{F_{\rm med}(z,R)}}{1+F_{\rm med}(z,R )}\right]  \Bigg\}\, .
\end{align}
The modification factor $F_{\rm med}$ should vanish at small angles, where one recovers that $\Delta S \ll 1$. However, once the angular scale becomes larger than the typical medium scale, then $\Delta S \gg0$. This suggests that the entropy might give a new handle on resolving QGP scales using jets.

\section{Conclusion}
In this work, we have computed the evolution of the single parton and jet density matrices in the presence of a QGP background.  Using these to compute the associated entropies, we show that this offer new interesting insights into the jet modifications due to evolution in the medium. Such quantities can be experimentally extracted and complete leading logarithmic computations will be provided in future work.

\section*{Acknowledgements}
 Y. M.-T. and J. B.'s work has been supported by the U.S. Department of Energy under Contract No.~DE-SC0012704. Y. M.-T. acknowledges support from the RHIC Physics Fellow Program of the RIKEN BNL Research Center.  J.-P. B. acknowledges partial support from the
European Union's Horizon 2020 research and innovation
program under grant agreement No 824093 (STRONG-
2020).\\

\bibliography{references.bib}

\begin{thebibliography}{3}

\bibitem{Barata:2023uoi}
J.~Barata, J.P. Blaizot, Y.~Mehtar-Tani, Phys. Rev. D \textbf{108}, 014039
  (2023), \texttt{2305.10476}

\bibitem{Neill:2018uqw}
D.~Neill, W.J. Waalewijn, Phys. Rev. Lett. \textbf{123}, 142001 (2019),
  \texttt{1811.01021}

\bibitem{Baier:2001yt}
R.~Baier, Y.L. Dokshitzer, A.H. Mueller, D.~Schiff, JHEP \textbf{09}, 033
  (2001), \texttt{hep-ph/0106347}

\end{thebibliography}

\end{document}